\documentclass{pazhastl}
\usepackage{mathptmx}
\usepackage{graphicx}
\usepackage{natbib}

\def\xss{XSS\,J00564$+$4548}
\def\igr{IGR\,J00234$+$6141}
\def\rxte{\emph{RXTE}}
\def\integral{\emph{INTEGRAL}}
\def\swift{\emph{SWIFT}}
\def\rosat{\emph{ROSAT}}

\begin{document}

\journalinfo{2005}{0}{0}{1}[0] 

\title{\xss\ and \igr\ --- new cataclysmic variables from \rxte\ and
  \integral\ all sky surveys}

\author{I.~F.~Bikmaev\email{ilfan.bikmaev@ksu.ru}\address{1,2}, 
  M.~G.~Revnivtsev\address{3,4},
  R.~A.~Burenin\address{3},
  R.~A.~Sunyaev\address{3,4}
  \addresstext{1}{Kazan State University, ul. Kremlevskaya 18, Kazan, Russia} 
  \addresstext{2}{Academy of Sciences of Tatarstan, ul. Baumana, 20, Kazan,
    Russia}
  \addresstext{3}{Space Research Institute (IKI), ul. Profsoyuznaya 84/32,
    Moscow, Russia}
  \addresstext{4}{Max-Planck-Institut f\"ur Astrophysik, Garching, Germany}
}

\shortauthor{Bikmaev \etal} 
 
\shorttitle{\xss\ and \igr\ --- new cataclysmic variables}

\submitted{\today}

\begin{abstract}
  
  We present the results of optical identification of two X-ray sources from
  \rxte\ and \integral\ all sky surveys: \xss\ and \igr. Using the optical
  data from Russian-Turkish 1.5-m Telescope (RTT150) and \swift\ X-ray
  observations, we show that these sources most probably are intermediate
  polars, i.e.\ binary systems with accreting white dwarfs with not very
  strong magnetic field ($\la$10\,MG). Periodical oscillations of optical
  emission with periods $\approx 480$ and $\approx 570$~s were found. We
  argue that these periods most probably correspond to the rotating periods
  of the white dwarfs in these systems. Further optical observations
  scheduled at RTT150 will allow to study the parameters of these systems in
  more detail.


\end{abstract}

\begin{figure*}
  \centering
  \centerline{
    \includegraphics[width=\columnwidth]{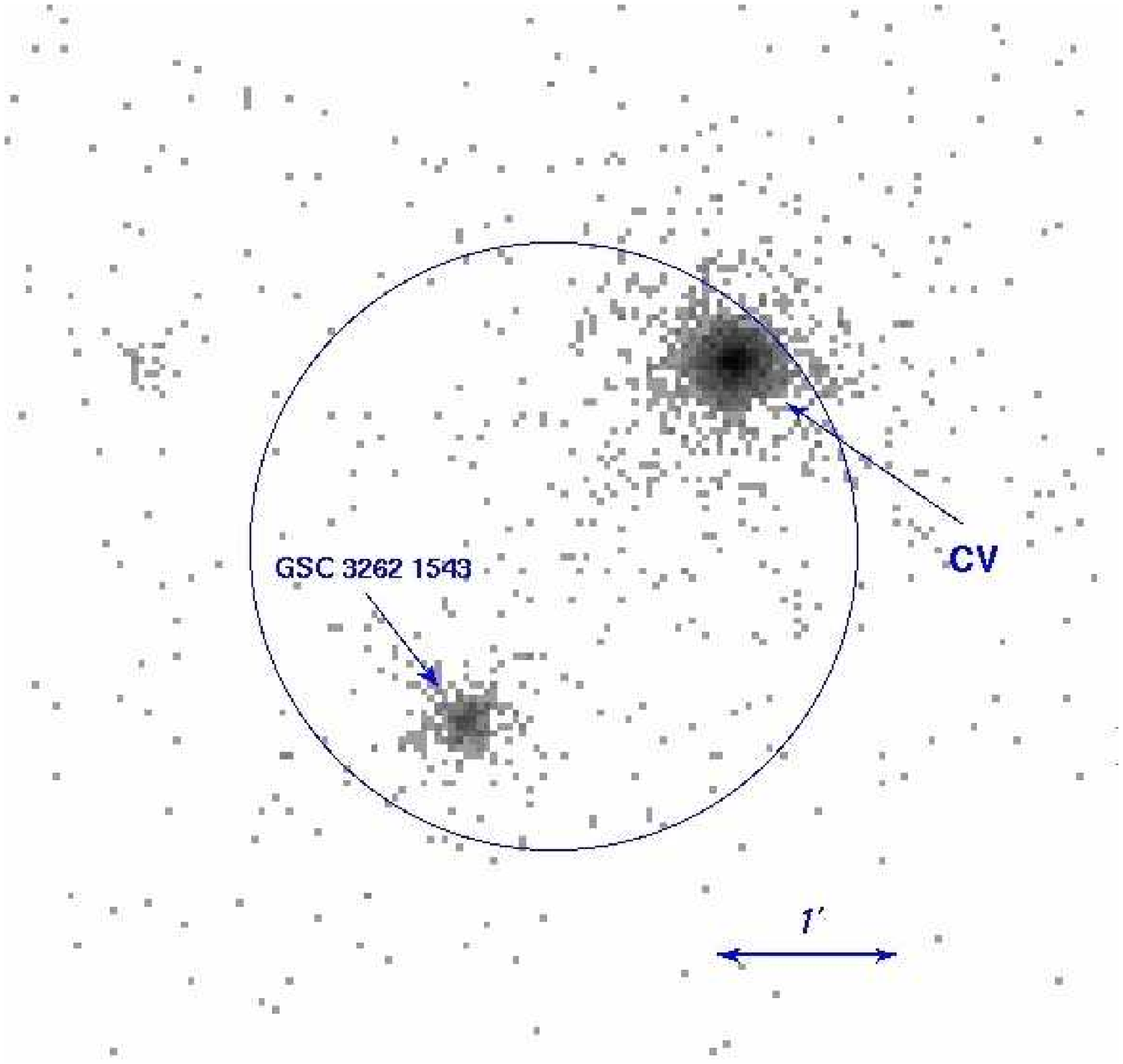}
    ~~~
    \includegraphics[width=\columnwidth]{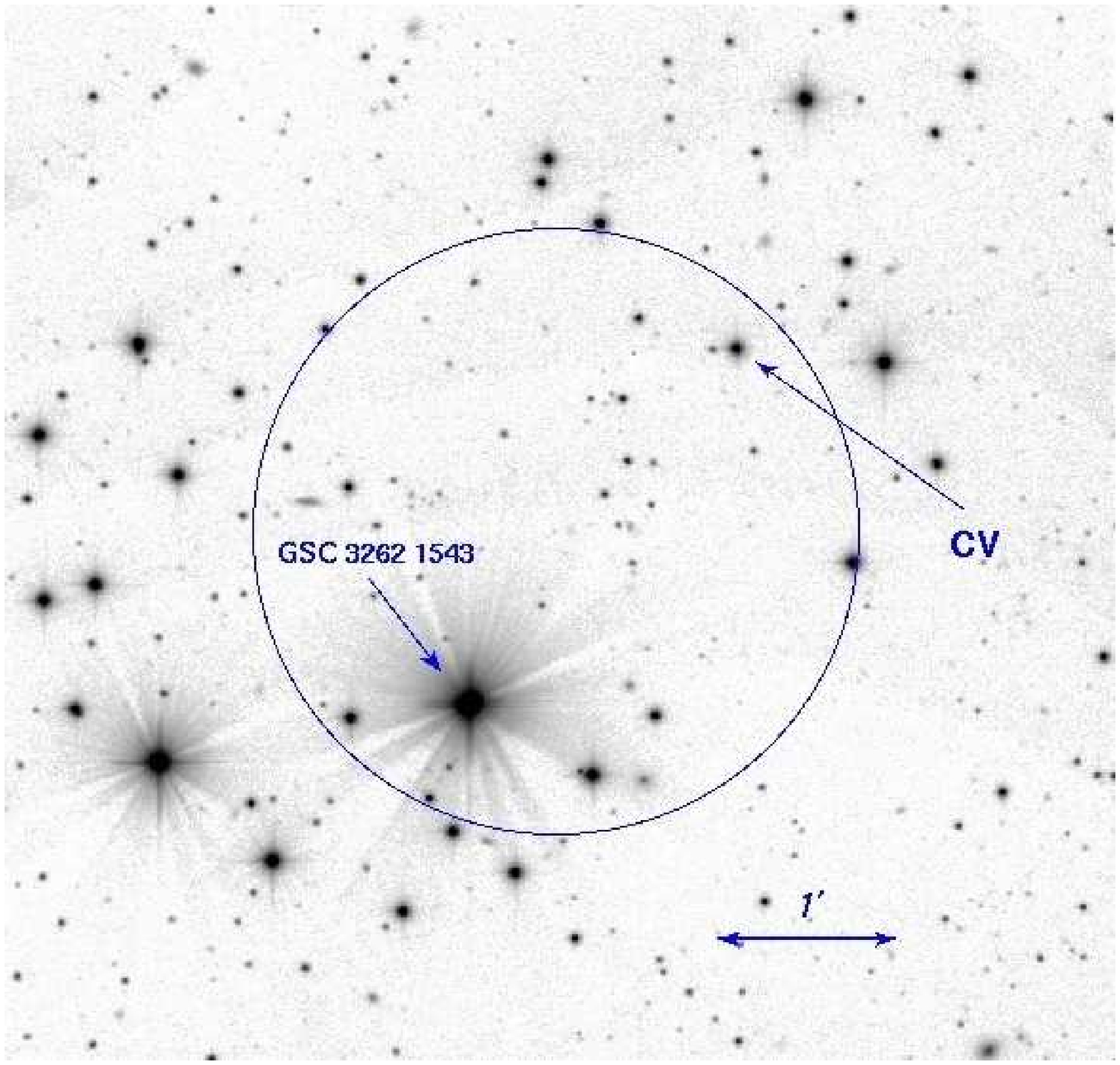}
  }
  \caption{The X-ray and optical images of the \xss\ field. Left ---
    \swift/XRT X-ray image, the error circle of \rosat\ source is also
    shown. Right --- RTT150 optical R-band image. In both images North is up
    and East is left, scale is similar and shown in the images.}
  \label{fig:fchart}
\end{figure*}

\section*{INTRODUCTION}

The most information on the population of low-luminous X-ray binaries comes
from \rosat\ All Sky Survey \citep[RASS, ][]{voges99}. The X-ray
luminosities of these low-luminous binaries are $\sim 10^{30\div33}$~erg/s
which corresponds to distances up to only one kpc in RASS and therefore are
not concentrated to Galactic disk. Deeper \rosat, \emph{Chandra} and
\emph{XMM-Newton} pointings can not provide more data because fainter
objects can only be found in Galactic plane where their surveyed volume is
much smaller and observations are hampered by Galactic absorption and large
number of faint galactic stars. Nearby objects are also more interesting
because they are bright and can be studied in more detail.

Unfortunately, for observations of some classes of X-ray sources RASS energy
band (0.2--2.4~keV) is far not optimal. For example, spectra of accreting
white dwarfs with $\sim 10^6$~G magnetic fields frequently have significant
intrinsic photoabsorption because of matter falling to their polar caps.
Soft X-ray emission ($<2$~keV) of these white dwarfs can be strongly
suppressed compared to harder X-ray emission ($>3$~keV).

The most sensitive all-sky X-ray survey completed to date in this harder
energy band is \rxte\ Slew Survey \citep[XSS,][]{mikej04}. This survey
contains useful information on the population of low-luminous X-ray
binaries. The numbers of these systems in XSS and RASS are actually
comparable, in spite of very different total number of X-ray sources
detected in these surveys. For example, the total number of known
intermediate polars is 44 \citep{rc03}, while the number of these objects
among identified XSS sources is 13 \citep{mikej04,ss05}. The reason is that
the main part of objects identified with bright stars in RASS are
chromospherically active stars with much softer X-ray spectrum which are not
detected in XSS.

Due to poor angular resolution of XSS the nature of 18 XSS sources (out of
294) remains unknown. According to general characteristics of unidentified
XSS objects, the most of them should be nearby active galactic nuclei
(AGNs), but there should be also Galactic binary systems, most of them ---
accreting white dwarfs \citep{mikej04}.

The total number of known accreting white dwarfs is not large
\citep[e.g.,][]{rc03} and discovery of any subsample of these objects gives
useful information to study of population of white dwarfs in Galaxy.  Even
more useful is to study accreting white dwarfs in statistically definite
survey. This allow to directly calculate volume density of these objects and
their contribution to Galactic X-ray emission \citep[see e.g.,][]{ss05}.
This is why the identifications of the sources discovered in \rxte\ and
\integral\ all-sky surveys are very important.

In 2005 we initiated a program of optical identifications of X-ray sources
from \rxte\ and \integral\ all-sky surveys. In frames of this program we
already identified six objects as previously unknown nearby AGNs
\citep{ilfan06}. In this paper we show that according to the results of
observations of \swift\ XRT in X-rays and Russian-Turkish 1.5-m Telescope
(RTT150) in optical, sources \xss\ and \igr\ are binary systems with
accreting white dwarfs, most probably intermediate polars.

\section*{Observations}
\label{sec:obs}

Optical data were obtained using two instruments at RTT150 telescope ---
CCD-photometer and spectrometer TFOSC. Short description of these
instruments and methods of data reduction, used in our work on optical
identifications of \rxte\ and \integral\ objects, are given in
\cite{ilfan06}.

\subsection{\xss}
\label{sec:xss}

The object \xss\ was found in \rxte\ all sky survey and identified with
X-ray \rosat\ source 1RXS\,J005528.0$+$461143 detected in RASS. The accuracy
of X-ray localization for this X-ray source ($\approx40\arcsec$) is
insufficient to identify this source with one of the optical sources in this
field.

\begin{figure}
  \centering
  \includegraphics[width=\columnwidth]{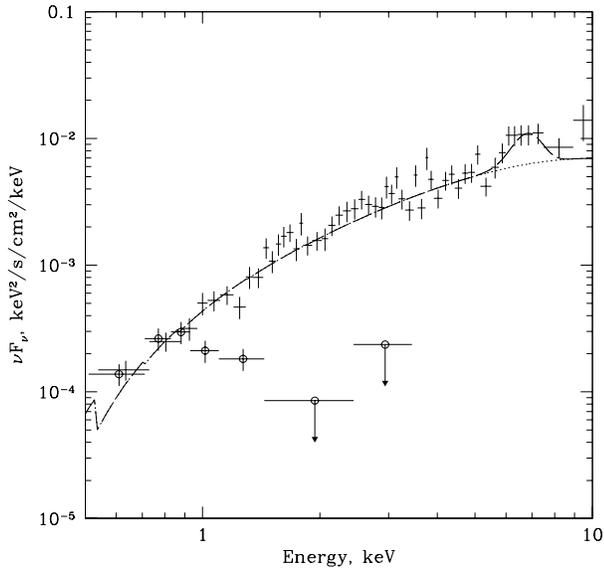}
  \caption{Spectra of two X-ray sources found by \swift/XRT in the error
    circle of \rosat\ source. The spectrum of star GSC\,3262\,1543 is shown
    with open circles, CV spectrum is shown with crosses.}
  \label{fig:swsp}
\end{figure}

During May -- October 2005 time period XRT telescope aboard \swift\
observatory \citep{swift} made four observations of this source. In result
of these observations it was found that there are actually two X-ray sources
in this field which were not resolved by \rosat, each of them giving
comparable contribution into \rosat\ source flux. The X-ray and optical
images of this field are shown in Fig.\,\ref{fig:fchart} and X-ray spectra
of these two sources are shown in Fig.\,\ref{fig:swsp}. One of the sources
is associated with star GSC\,3262\,1543, $V\approx11^m$. Its X-ray spectrum
resemble that of late spectral class star with chromospheric activity or
binary system of RS CVn type \citep[see e.g., ][]{schmitt90}. This source is
soft and could not be detected in XSS, while second X-ray source is much
harder (see Fig.\,\ref{fig:swsp}). Therefore, we conclude that \xss\ should
be identified with that second \swift\ X-ray source and with optical object
at coordinates $\alpha=00^h 55^m 20^s.0$, $\delta=+46^\circ 12\arcmin
57\arcsec$ (J2000).

The X-ray spectrum of this source can be adequately described by a simple
power law with photoabsorption with photon index $\Gamma\sim0.8\pm0.1$ and
photoabsorption parameter $N_HL=(0.22\pm0.05)\times 10^{22}$~cm$^{-2}$.
Galactic extinction from radio 21~cm data is $0.11\times 10^{22}$~cm$^{-2}$
\citep{dikeylockman90}, i.e.\ somewhat smaller than that seen in X-ray
spectrum, which might indicate the presense of intrinsic photoabsorption in
the source emission. Photon index is not typical for extragalactic objects
(AGNs), which can be expected on that high Galactic latitude, while is very
similar to typical photon indexes of accreting white dwarfs --- cataclysmic
variables (CV).

At the detection threshold in this spectrum there is also a wide emission
line near $\approx6.7$~keV, which is also typical for CVs and not typical
for AGNs. The line is detected with $\Delta \chi^2$ corresponding to 99.9\%
probability that this feature is not accidental deviations of the measured
points from the power law fit. The equivalent width of this wide line is
$EW=1.06\pm0.30$~keV which is also expected for optically thin plasma
emission from accreting white dwarf.

\begin{figure}
  \bigskip
  \medskip
  \includegraphics[width=\columnwidth]
  {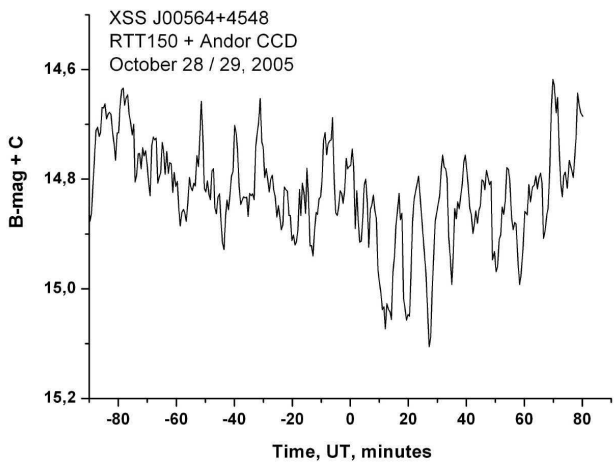}
\caption{\xss\ light curve obtained with RTT150 on Oct.\ 28 2005.}
\label{lcurve}
\end{figure}

\begin{figure}
  \includegraphics[width=\columnwidth]{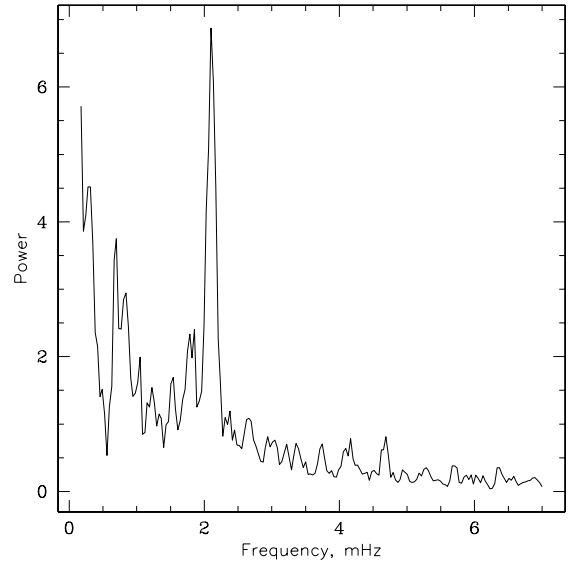}
  \caption{Power spectrum of optical variability of \xss. The prominent peak
    on $(480~s)^{-1}$ is clearly seen.}
  \label{lomb}
\end{figure}

\begin{figure}
  \includegraphics[width=\columnwidth]{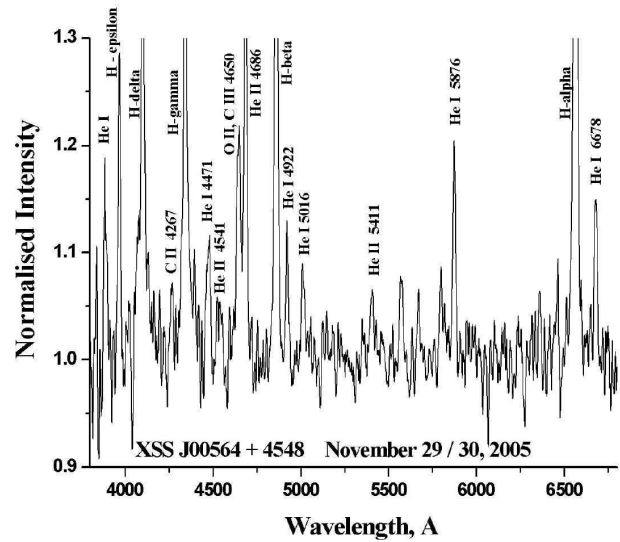}
  \caption{Optical spectrum of \xss, obtained with RTT150 on 29 Nov.\ 2005.}
  \label{opt_spectrum}
\end{figure}

\begin{figure}
  \includegraphics[width=\columnwidth]{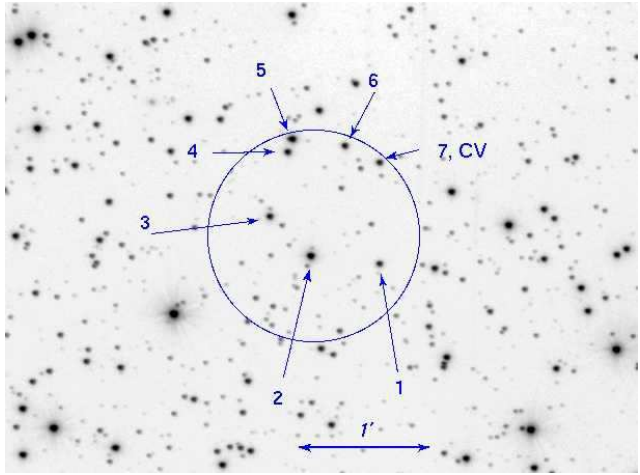}
  \caption{Optical image of the field of \igr. The numbers show sources for
    which optical spectra were obtained. Orientation is North-up,
    East-left, scale is shown in the image.}
  \label{igr_pos}
\end{figure}

\begin{figure}
  \includegraphics[width=\columnwidth]{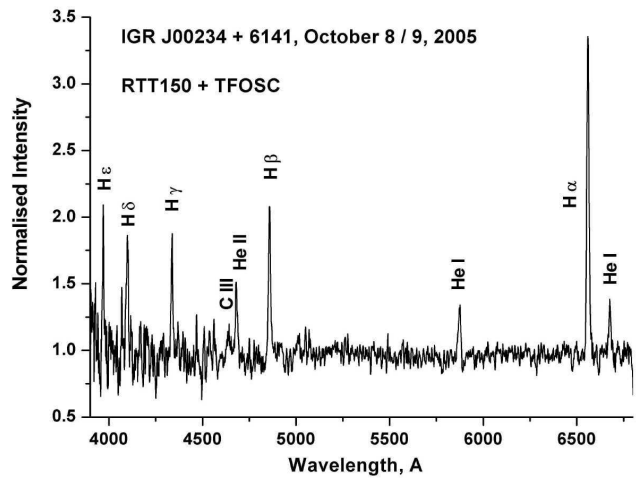}
  \caption{Optical spectrum of \igr, obtained with RTT150 on Oct.\ 8 2005.}
  \label{igr_opt_spectrum}
\end{figure}

\begin{figure}
  \includegraphics[width=\columnwidth]{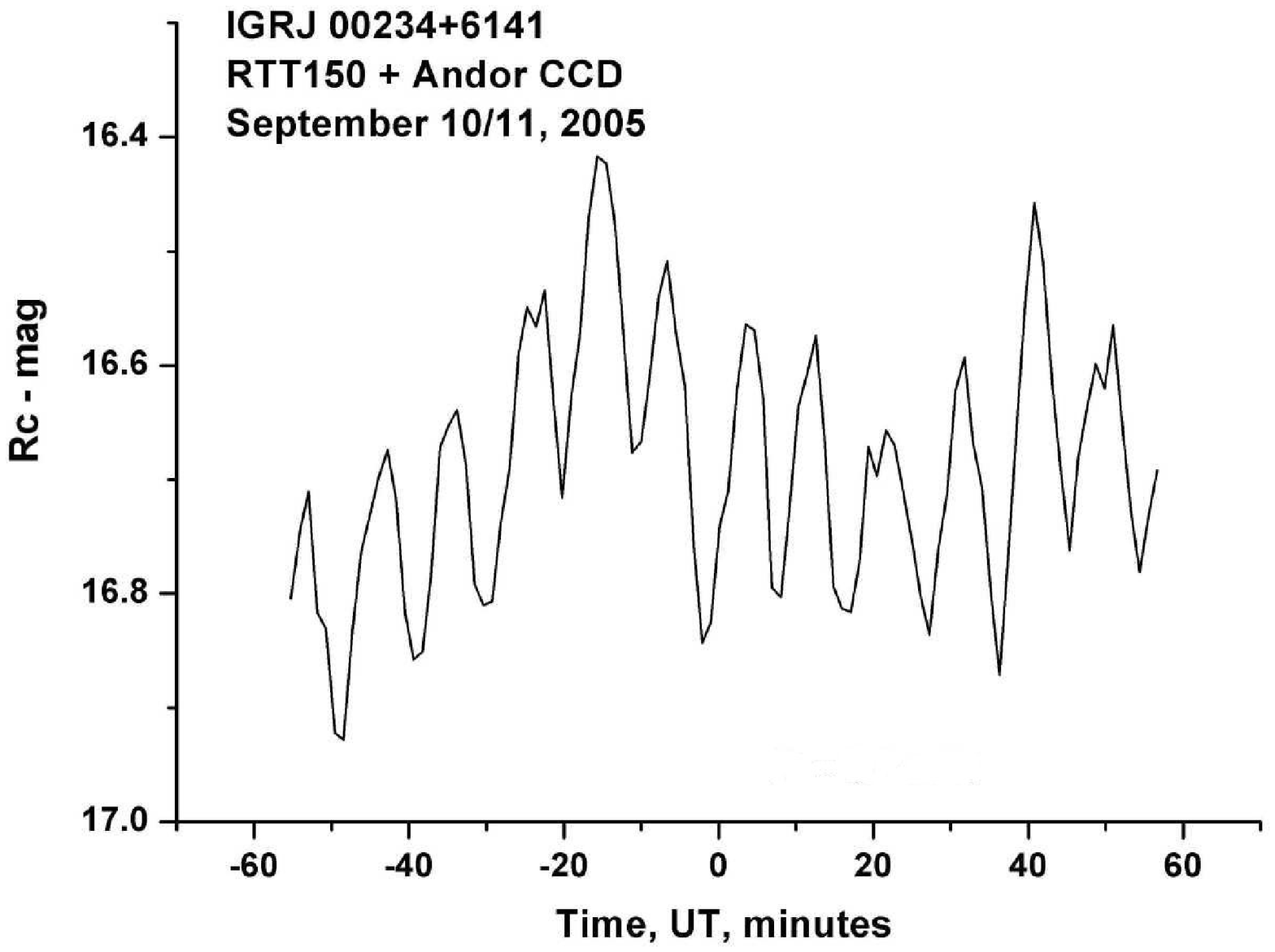}
  \caption{The optical light curve of \igr\ in $R_c$, obtained with RTT150 on
    Sep.\ 10 2005.}
  \label{igr_lc}
\end{figure}

In order to check the hypothesis on the nature of object \xss, we made a set
of otical photometrical observations on Aug.\ 18, 20, Sep.\ 11 and Oct.\ 19,
20, 21, 22, 23, 26 and 28 2005. In Fig.\,\ref{lcurve} the light curve of the
optical source obtained on Oct.\ 28 is shown as an example. We detected
strong variability of the optical source with amplitude $\Delta B\approx
0.3$ on the time scales down to $\sim$50--100~s, which completely exclude a
possibility that this source may be an AGN. A subsequent analysis showed
that this variability contains very significant periodic component. In
Fig.\,\ref{lomb} we show a Lomb-Scargle periodigram \citep{lomb76} averaged
over all available data.  The prominent peak on $(\approx 480~s)^{-1}$ is
clearly seen in this figure. The measured period is most probably not an
orbital period of binary system, which is typically an order of magnitude
longer, but a rotation period of white dwarf.

We emphasize, that we have no direct information that the period of otpical
variability is correspondent to just the rotation period of white dwarf and
not to the orbital period of this system. However, there are only few CV
with orbital periods $<1$~hour are known to date \citep{rc03}. The most of
them are AM~CVn type variables --- systems with no hydrogen lines in their
optical spectra, since in these systems the accretting matter comes from a
degenerate star. Optical spectrum obtained with RTT150 telescope is shown in
Fig.\,\ref{opt_spectrum}. The spectrum show bright emission lines, including
hydrigen ones, typical for accreting white dwarfs \citep[see
e.g.,][]{williams82,schmidt86}. Therefore, in this system the accretion
should most probably be from a normal star. These systems usually have
orbital periods of order of few hours \citep{rc03}.

The magnetic field of white dwarf in this system should be strong enough in
order to crease assymetry in the accretion flow, which we see as pulsations.
On the other hand its magnitude should not be larger than $\sim10$~MG,
because in this case the magnatic field of the white dwarf can lock on the
secondary star and will synchronize the orbital and wite dwarf rotational
periods \citep[see e.g.][]{patterson94}. Therefore, on the basis of optical
and X-ray data we can classify this object as intermediate polar, i.e.\
accreting white dwarf with not very strong magnetic field ($\la$10~MG).

\subsection{\igr}

X-ray source \igr\ was discovered in a deep \integral\ observations of
supernova remnant Cas A \citep{hartog05}. The position of the source
(accuracy $\approx 3\arcmin$) is consistent with \rosat\ source
1RXS\,J002258.3$+$61411 with much better localization ($\approx 10\arcsec$).
This good localization accuracy allowed to search the optical companion of
this source between only few optical objects.

In Oct.\ 2005 we obtained spectra of seven most bright optical objects in
\rosat\ error circle. The finding chart of the \xss\ field is shown in
Fig.\,\ref{igr_pos}. The spectrum of seventh object showed strong emission
lines with zero redshift, which is the evidence that this is the Galactic
object. The spectrum of this source is shown in
Fig.\,\ref{igr_opt_spectrum}. A set of emission lines is very typical for
cataclysmic variables --- accreting white dwarfs \citep[see e.g.,
][]{williams82,schmidt86}. The X-ray ray source \igr\ should be identified
with this CV, since the probability of chance finding of any CV in \rosat\
error circle is negligible. The coordinates of this object are $\alpha=00^h
22^m 57^s.6$, $\delta=+61^\circ 41\arcmin 08\arcsec$ (J2000).

In Jan.\ 2006 this object was independently observed by other group
\citep{halpern06}.  From their obtained optical spectrum these authors are
also found that this object is a CV and that X-ray source \igr\ should
identified with it.

Optical photometric light curves obtained on Sep.\ 10 2005 showed the
presence of prominent periodicity with period 570~s. A part of the light
curve is shown in Fig.\,\ref{igr_lc} as an example. Periodical oscillations
are easily seen in this light curve even with naked eye. Like in case of
\xss\ (see above), due to small value of measured optical period we propose
that if correspond to the rotation period of white dwarf. Therefore, like in
previous case, we can classify this object as an accreting weakly magnetized
white dwarf --- intermediate polar.

\section*{Conclusions}

Using observations of RTT150 in optical and \swift\ observatory in X-rays,
we identified two \rxte\ and \integral\ X-ray sources from their all-sky
surveys \xss\ and \igr. Obtained photometric light curves and optical
spectra propose that these objects most probably are accreting binary
systems with magnetized white dwarfs --- intermediate polars.

\akcnowledgements

Authors are grateful to Hans Ritter and anonymous referees for valuable
comments. Authors are also grateful to the director of T\"UBITAK National
observatory (Turkey) Prof.\ Zeki Aslan for his strong support of this work,
to A.~Galeev, R.~Zhuckov (KGU), D.~Denisenko and A.~Mescheryakov (IKI) for
their help in performing optical observations.  This work was supported by
grants RFFI~05-02-17744, RFFI~05-02-16540, NSH-1789-2003.2 and
NSH-1100.2006.2, and also by program of Presidium of the Russian Academy of
Sciences ``Formation and evolution of stars and galaxies'', and by DFG
Priority Programme 1177 ``Witnesses of Cosmic History: Formation and
evolution of galaxies, black holes, and their environment''. RB also
acknowledge support by grant of President of RF, MK-4064.2005.2.

\end{document}